\documentstyle[twoside,fleqn,epsf,espcrc2]{article}
\title{Quark Confinement and Surface Critical Phenomena 
     \thanks{Presented by J.~Kuti at Lattice 99, Pisa, June 29 - July 3, 1999.
     Supported by DOE Grant DE-FG03-97ER40546.\newline
     UCSD/PTH/99-19, FERMILAB-Conf-99/279-T
}}

\author
{K.J.~Juge\address{Fermi National Accelerator Laboratory, 
  P.O.~Box 500, Batavia, Illinois 60510}, 
  J.~Kuti\address{Dept.~of Physics, University of California at San Diego,  
  La Jolla, California 92093-0319} 
  and C.J.~Morningstar${}^{\rm b}$}

\begin{document}

\begin{abstract}
Surface critical phenomena and the related onset of 
Goldstone modes probe the fundamental properties of the confining flux
in Quantum Chromodynamics. New ideas on surface roughening and their
implications for lattice studies of quark confinement 
are presented. Problems with the oversimplified string description of the 
Wilson flux sheet are discussed.
\end{abstract}

\maketitle

\section{The spectrum of the confining flux}

\begin {figure}[h]
\begin{center}
\epsfxsize=3.0in\epsfbox[25 52 545 575]{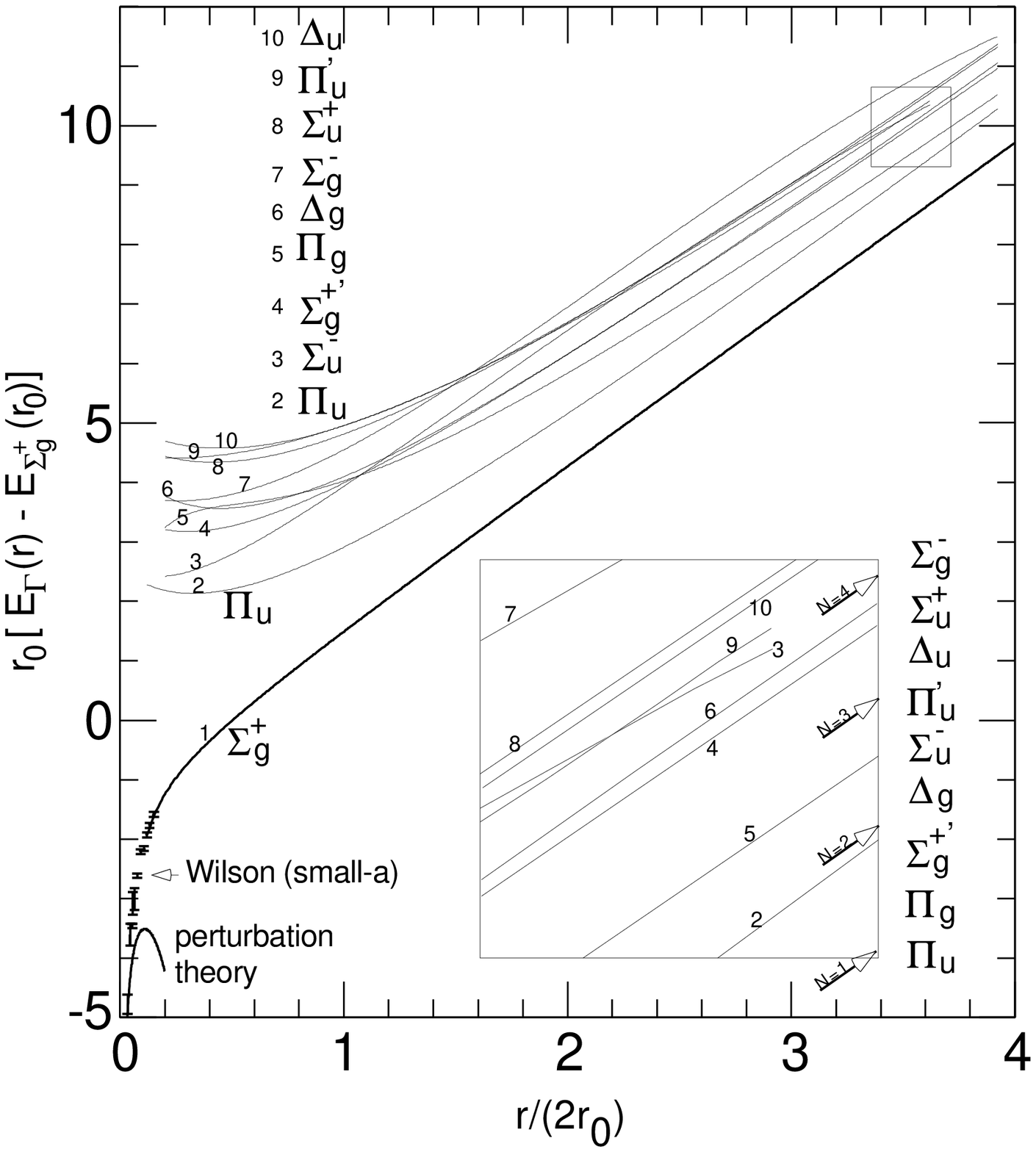}
\end{center}
\vskip -0.6in
\caption{Continuum limit extrapolations are shown for
$r_0 [E_\Gamma(r)-E_{\Sigma_g^+}(r_0)]$ against $r/(2r_0)$ for
ten different states characterized by quantum numbers $\Gamma$ .  
The arrows in the inset show the locations of the four lowest string
modes and the associated discrepancies in the spectrum 
(for notations and further discussions, see Refs.~\cite{JKM1,JKM2}).}
\label{fig:spectrum}
\end{figure}

A rather comprehensive determination of the rich energy spectrum 
of the confined chromoelectric flux 
between static sources in the fundamental representation of ${\rm
SU(3)_c}$ was reported earlier~\cite{JKM1,JKM2}
for separations r ranging from 0.1 fm to 4 fm. The full spectrum
is summarized in Fig.~\ref{fig:spectrum} with different
characteristic behavior on two scales separated 
approximately at r=2 fm.

\subsection*{No string spectrum for r ${\mathbf \leq 2}$~fm}

The $\Sigma_{\rm g}^+$ ground state is the familiar static 
quark-antiquark potential which is dominated by  
the rather dramatic linearly-rising behavior
once ${\rm r}$ exceeds about ${\rm 0.5~fm}$. 
Although the empirical function
${\rm E}_{\Sigma_{\rm g}^+}({\rm r}) = -{\rm c}/{\rm r} + \sigma{\rm r}$
approximates the ground state energy very well for 
${\rm r}\geq {\rm 0.1~fm}$, the fitted constant ${\rm c = 0.3}$
has no relation to the running Coulomb law whose loop expansion
breaks down before ${\rm r=0.1~fm}$ separation is reached~\cite{peter}. 
Early indoctrination on 
the popular string interpretation of the confined flux
for ${\rm r}\leq{\rm 2~fm}$ was mostly based on the observed shape of the 
$\Sigma_{\rm g}^+$ ground state energy: the
linear shape of the ground state
potential for ${\rm r} \geq~{\rm 0.5~fm}$ and the approximate agreement
of the curvature shape for ${\rm r} \leq 0.5~{\rm fm}$ with the ground
state Casimir energy $-\pi/(12{\rm r})$ of a long 
confined flux~\cite{Luescher2}.
The excitation  spectrum clearly contradicts 
earlier claims~\cite{michael} on 
the simple string interpretation of the linearly rising confining 
potential.
The gluon excitation energies lie well below the string predictions
and the level orderings and degeneracies
are in violent disagreement with expectations from a fluctuating string.

\subsection*{Goldstone modes for r ${\mathbf > 2}$~fm}

A feature of any low-energy description of a fluctuating
flux sheet in euclidean space is the presence of 
Goldstone excitations associated with the 
spontaneously-broken transverse translational symmetry.  These transverse
modes have energy separations above the ground state given by multiples of
$\pi/r$ (for fixed ends).  The level orderings and approximate degeneracies
of the gluon energies at large r match, without exception, those expected
of the Goldstone modes.  However, the 
precise $\pi/{\rm r}$ gap behaviour is not
observed. The spectrum is consistent with massive capillary waves
on the surface of the flux sheet, with a cutoff dependent mass gap.
The most likely explanation for this gap is Peierls-Nabarro lattice 
pinning
of the confining flux sheet at small correlation lengths.
Our new results~\cite{JKM3} on the same spectrum in SU(2) for D=3,4, 
and a detailed
test of the strong coupling spectrum in SU(2) for D=3 lend further
support to the above summary of the earlier findings.

\section{Z(2) model at D=3}

\begin {figure}[h]
\begin{center}
\epsfxsize=3.0in\epsfbox[50 50 554 770]{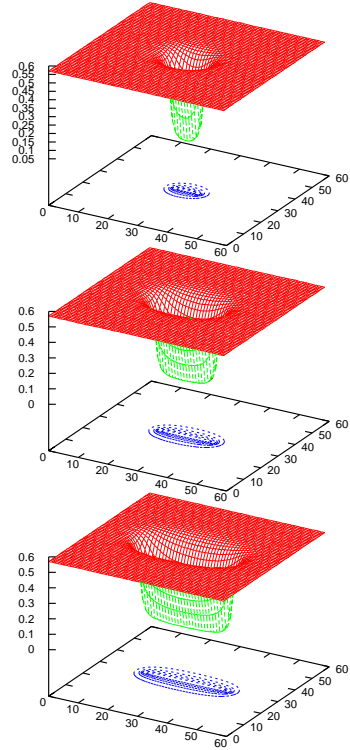}
\end{center}
\vskip -0.6in
\caption{Flux formation at corr. length 
$\xi \approx 2$ for spatial Wilson loop
sizes r=10,20,30.}
\label{fig:flux}
\end{figure}

\begin {figure}[h]
\begin{center}
\epsfxsize=3.0in\epsfbox[50 50 554 770]{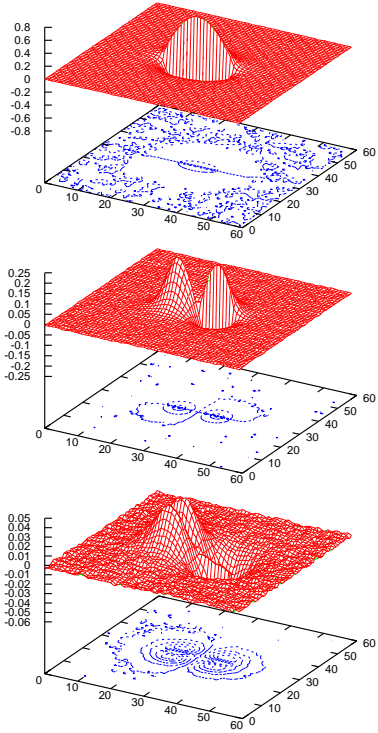}
\end{center}
\vskip -0.6in
\caption{Wavefunctions of the excited Z(2) flux at 
spatial Wilson loop size r=20 for $\xi \approx 2$. 
The first two Goldstone modes
and a massive intrinsic excitation are shown.}
\label{fig:wf}
\end{figure}

\begin {figure}[h]
\begin{center}
\epsfxsize=2.5in\epsfbox[18 144 592 718]{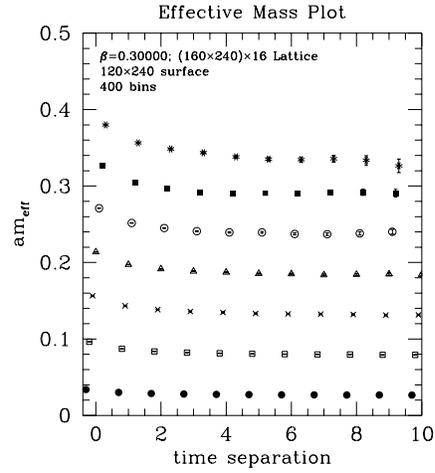}
\end{center}
\vskip -0.6in
\caption{Effective mass plot of the string spectrum for a very large
Wilson loop at $\xi \approx 1/2$ before the Peierls-Nabarro gap opens up.}
\label{fig:string}
\end{figure}

\begin {figure}[h]
\begin{center}
\epsfxsize=2.5in\epsfbox[0 0 600 800]{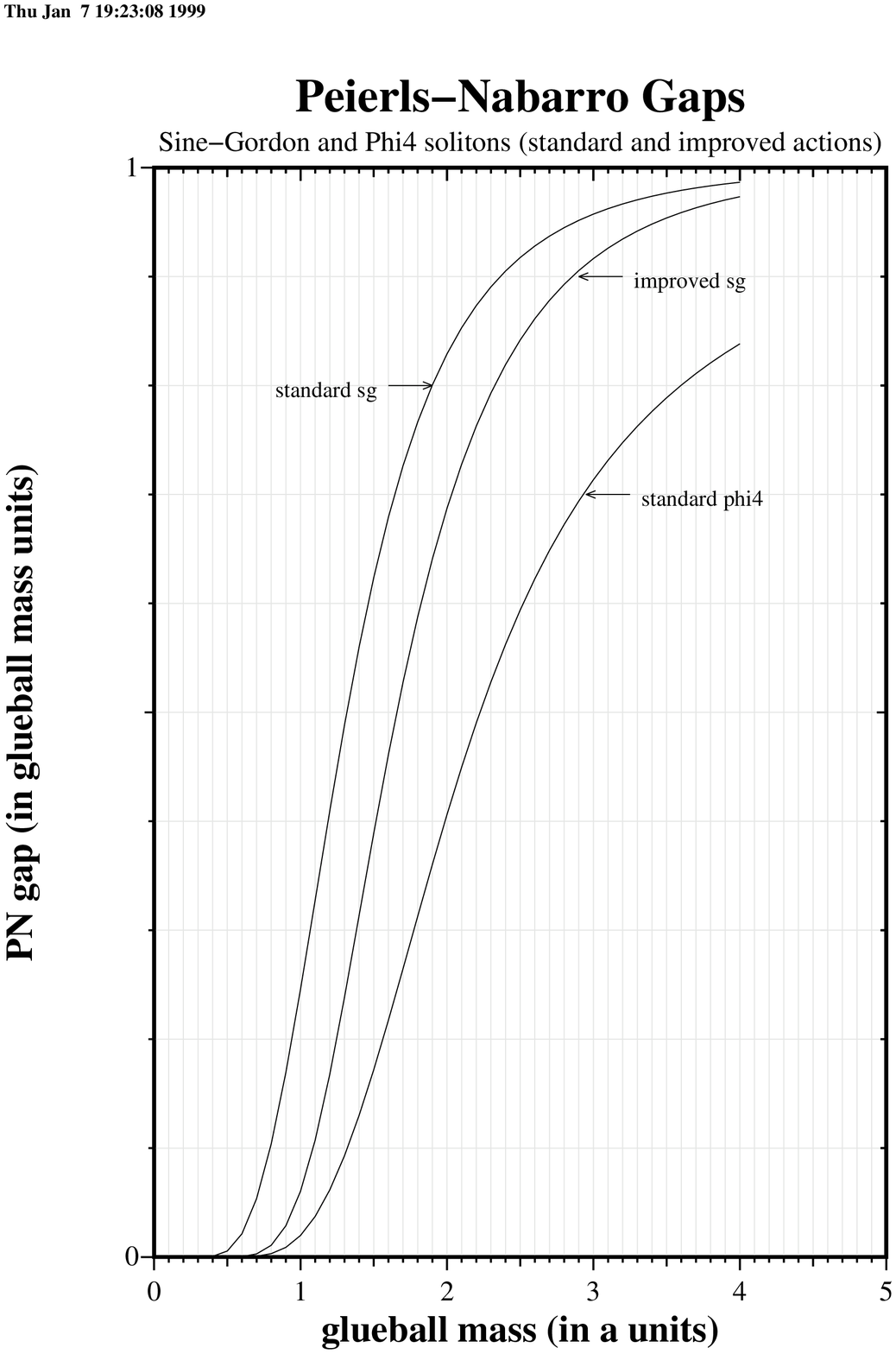}
\end{center}
\vskip -0.6in
\caption{The Peierls-Nabarro gap (negligible at large
correlation lengths, or inverse glueball mass) implies pinning and mass
generation for the Goldstone modes.}
\label{fig:peierls}
\end{figure}

The purpose of the Z(2) project is to understand flux formation and the 
string excitation spectrum from high statistics
simulations and the loop expansion on the analytic side.
The WKB approximation of flux formation in  the sine-Gordon field  
representation of the monopole plasma was discussed earlier~\cite{JK}.
The Z(2) model maps into the Ising model by duality transformation
which was exploited before in the study of large Wilson loops~\cite{Gliozzi}.
By invoking universality, the critical region of the Z(2) model is mapped
into D=3 $\Phi^4$ scalar field theory in the study of flux formation.
The confining flux sheet of the Wilson loop corresponds to a twisted
surface in the Ising representation which is described by a classical
soliton solution of the $\Phi^4$ field equations.
Excitations of the flux are given by
the spectrum of the fluctuation operator 
$
{\mathcal M} = -\nabla^2 + {\rm U}^{\prime\prime}(\Phi_{\rm soliton})
$
where ${\rm U(\Phi)}$ is the field potential energy of the $\Phi^4$ field.
The spectrum of the fluctuation operator ${\mathcal M}$ of the
finite surface is determined
from a two-dimensional Schr\"odinger equation with a
potential of finite extent~\cite{JK}. 
In the limit of asymptotically large surfaces,
the equation becomes separable in the longitudinal and transverse
coordinates.
The transverse part of the spectrum is in close analogy with the 
quantization of the one-dimensional classical $\Phi^4$ soliton.
There is always a discrete zero mode in the spectrum which is 
enforced by translational invariance in the transverse direction. 
Figs.~\ref{fig:flux},\ref{fig:wf},\ref{fig:string},\ref{fig:peierls}
illustrate some of the results which are 
{\em consistent with our findings in QCD}.

\section*{Acknowledgements}
One of us (J.~K.) would like to acknowledge valuable discussions 
with S.~Renn, P.~Hasenfratz, and F.~Niedermayer
on surface Goldstone modes.

\end{document}